# High-Fidelity Single-Pixel Imaging at Ultra-Low Sampling Ratios via Physically Enhanced Laguerre Gaussian Encoding

JunYi Xiong[1, †], Can Su[1, †], ZhiYuan Wang[2], YangYang Xiong[3], HongJie Wang[1], MengQiang Cai[4,5], GuiYuan Cao[6,7], XioaHua Deng[4,5], ZhongQuan Nie[8], WeiChao Yan[4,5], BaoHua Jia[7]

[1] Department of Physics, School of Physics and Materials Science, Nanchang University, Nanchang 330031, China;

[2] Department of Computer Sciences, School of Mathematics and Computer Sciences, Nanchang University, Nanchang 330031, China;

[3] Department of Electronic and Information Engineering, School of Information and Engineering, Nanchang University, Nanchang 330031, China;

[4] Institute of Space Science and Technology, Nanchang University, Nanchang 330031, China;

[5] Engineering Research Center of Intelligent Sensing Technology in Space Information, Ministry of Education, Nanchang University, Nanchang 330031, China;

[6] Nanophotonics Research Center, Institute of Microscale Optoelectronics and State Key Laboratory of Radio Frequency Heterogeneous, Shenzhen University, Shenzhen 518060, China;

[7] Centre for Atomaterials and Nanomanufacturing (CAN), School of Science, RMIT University, Melbourne, VIC 3000, Australia;

[8] College of Advanced Interdisciplinary Studies, National University of Defense Technology, Changsha 410073, China.

† These authors contributed equally to this work.

Corresponding author: ZhongQuan Nie (e-mail: niezhongquan1018@163.com), WeiChao Yan (e-mail: yanweichao@ncu.edu.cn), BaoHua Jia (e-mail: baohua.jia@rmit.edu.au)

This work was supported by the National Natural Science Foundation of China (12004155, 12364043, 62205214); Jiangxi Provincial Natural Science Foundation (20224BAB211016, 20224BAB211017, 20242BAB25102).

**Abstract**: Single-pixel imaging (SPI) has offered an unprecedented technique for capturing a targeted scenes without requiring either raster-scanned systems or muti-pixel detectors. However, in the current research, there are rare study reports about achieving both high spatial quality and low sampling ratio below 5% without additional algorithms in the existing SPI architectures. To circumvent these challenges, here we demonstrate a novel Laguerre Gaussian single-pixel imaging (LGSI) technique achieving ultra-low sampling ratio (3%) and super-high spatial imaging quality (Structural Similarity (SSIM) of 0.739 and a peak signal-to-noise ratio (PSNR) of 20.762 dB). The fundamental methodology relies on the enhancement of the encoded patterns by the differential modulation of discrete orthogonal physical LG moments, enabling the reconstruction of illuminated target object via a linear weighting of the structured light intensity. Leveraging this orthogonal mechanism, LGSI demonstrates superior imaging quality and computational efficiency, surpassing the capabilities of non-orthogonal moments. Comparative analyses of the power spectra from reconstructed images highlight the enhanced efficacy of LGSI over Hadamard SPI (HSI) and Fourier SPI (FSI). Our results suggest the possibility of encoding multi-dimensional structured light fields as a promising pathway for realizing low sampling ratio, universal, and physical-endow SPI.

# Introduction

Single-pixel imaging (SPI)[1-3] is a computational imaging technique that reconstructs images of a target scene utilizing a single-point detector and spatial light modulation, eliminating the need pixelated detectors typically employed in conventional array camera system. SPI produces images by projecting a series of structured illumination patterns and measuring the correlated intensity on a detector without spatial resolution. In essence, the spatial information of the object can be obtained solely through the utilization of post-coded patterns and the corresponding sequential bucket signals acquired by a single pixel detector with the object image subsequently reconstructed by computational algorithms. In contrast to array detectors such as CCD or CMOS, single-pixel detectors, including avalanche diodes and photo multiplier tubes, demonstrate enhanced detection sensitivity under low-light conditions[4]. Moreover, the SPI technique offers significant potential for application in non-visible[5-8], multispectral imaging[9, 10], remote sensing imaging[11], imaging through scattering media[12, 13] and optical encryption[14, 15]. However, SPI faces several challenges, including the difficulty of achieving crosstalk-free, high-quality imaging at low sample ratios. Developing of new methods to simultaneously address these challenges is imminent.

While SPI offers distinct advantages in computational imaging, it necessitates the use of a considerable number of illumination patterns for modulation. Early implementation of SPI employed random patterns[16] necessitating extensive oversampling to achieve acceptable image quality. However, the redundancy inherent in random patterns often results in poor imaging quality and reduced imaging speed. To enhance these limitations simultaneously, SPI has developed methods that are broadly divided into two primary approaches: algorithmic and physical methods.

In the context of algorithmic methods of SPI, compressive Sensing (CS)[3, 27-29] theory has been introduced to SPI with the aim of reducing the total number of measurements required, thereby enabling the reconstruction of the object image in circumstances where sampling is incomplete. Nevertheless, achieving higher compression ratios without compromising image quality remains, as the sparsity of coefficients diminishes, necessitating a trade-off between speed and quality. The incorporation of deep learning into SPI[34-37] has shown promise in reducing sampling ratio, but challenges persist due to the lack of inherent physical interpretability to deep learning, the requirement of a large dataset for training, and the instability in image reconstruction caused by the variability in network parameters across training iterations.

In contrast, physical methods in SPI avoid the uncertainties associated with iterative algorithms by introducing physical mechanisms for generating the basis of SPI. These methods efficiently utilize light-intensity information with orthogonal moments[30] being widely applied for high efficiency image reconstruct. Notable examples include Hadamard SPI (HSI)[13, 17-19] and Fourier SPI (FSI)[20-22]. However, cumbersome HSI needs to employ sorting algorithms[18, 31, 32], which reduce the sampling ratio in terms of compression perception. Furthermore, the infinite bandwidth property of the step function can result in the degradation of reconstructed images in finite diffraction systems[26, 33]. Similarly, the use of the FSI method necessitates the introduction of a phase shift, which imposes constraints on imaging speed.

Physically enhanced orthogonal moments[23-26] have great potential in advancing SPI technology, representing a key focus and challenge in the field. However, research in this area is still in the ascendant. Here, we propose a novel coding approach based on physically orthogonal Laguerre Gaussian (LG) moments calculated from LG function. The encoding of basis pattern is implemented in two dimensions in SPI imaging systems, enabling the reconstruction of object images from known LG patterns and the correlated intensity obtained from a single pixel detector. The results of the simulations and experiments consistently demonstrate that the proposed LG-based SPI (LGSI) scheme is capable of reconstructing images of high spatial quality, utilizing a limited number of patterns. LGSI offers exceptional image quality and superior imaging efficiency at extremely low sampling ratios without requiring supplementary processing or complex configurations. Moreover, LGSI does not necessitate supplementary processing or complex configuration, and can be readily extended to a vast array of SPI imaging paradigms, offering a versatile and efficient solution for high-quality imaging applications.

## Principles and Methods based on LGSI

The LG function of order $m$ and $n$ is given by

$$v_{mn}(r,\varphi) = C_{mn}\left(\sqrt{2}\frac{r}{\omega_{0s}}\right)^{|m|} L_n^{|m|}\left(2\frac{r^2}{\omega_{0s}^2}\right)\exp\left(-\frac{r^2}{\omega_{0s}^2}+im\varphi\right) \quad (1)$$

where is defined on polar coordinates of $r$ and $\varphi$, $L_n^m(\xi)$ is associated Laguerre polynomials, $m = 0, \pm 1, \pm 2, \ldots, \pm M, n = 0, \pm 1, \pm 2, \ldots, N, M, N > 0$, $\omega_{0s}$ is the waist radius of the fundamental mode of the Gaussian beam. $C_{mn}$ is normalization factor, and it is equal to

$$C_{mn} = \frac{1}{\sqrt{\iint |v_{mn}(r,\varphi)|^2 drd\varphi}} \quad (2)$$

A single $v_{mn}$ can generate two distinct patterns, representing the real and imaginary parts, respectively. Subsequently, two sets of patterns can be derived from the segmentation of positive and negative distributions within LG functions to adapt the DMD. Additionally, the patterns are resized and standardized to ensure compatibility with SPI reconstruction (see Supplementary information Section 1).

The sequence of patterns observed in LGSI imaging is contingent upon the ordering of $m$, $n$. It is established that the low-frequency information of low-order patterns has a greater weight value in the image than the high-frequency information of high-order patterns. The image is reconstructed using LG patterns with different sampling ratios, and reconstruction quality of the image is evaluated using peak signal-to-noise ratio (PSNR) and structural similarity (SSIM).

To evaluate the quality of LGSI, a typical image was reconstructed at varying sampling ratios, with the imaging resolution set to $128 \times 128$ pixels. The sampling ratio is defined as the proportion of the number of illumination patterns utilized to the total number of pixels in the target image. The imaging efficacy of LGSI is illustrated in Fig. 1. As shown in Table 1, the PSNRs begin to decrease after a sampling ratio of 6%, while the SSIMs increase up to sampling ratio of 20%, beyond which decreases with the characteristic noise of LG patterns appearing gradually. The limited number of pixels, constrained by the factorial term in the LG polynomial, results in a loss of accuracy for discrete high-order LG moments demonstrate.

Specifically, the values of these patterns become increasingly concentrated at larger radius distance within the image, leading to diminished reconstruction fidelity.

The number of LG patterns is jointly determined by the two parameters ($M$, $N$), allowing for a multitude of potential values at a given sampling ratio. For instance, setting ($M$, $N$) to (20, 20) yields the same sampling ratio as configurations such as (40, 10), (80, 5), and so forth. It is therefore necessary to consider the impact of different values of ($M$, $N$) on the image reconstruction results of LGSI in order to identify optimal parameter combinations for specific sampling ratios.

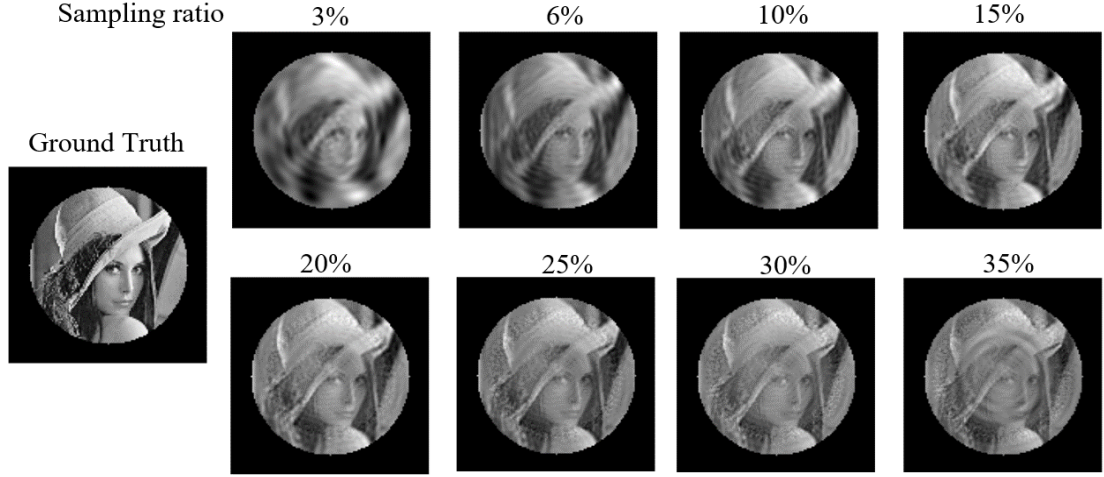

Fig. 1. The reconstructed results of LG method at different sampling ratios

Table 1. PSNR & SSIM evaluations for LGSI

| Sample ratio | PSNR (dB) | SSIM |
|---|---|---|
| 3% | 20.762 | 0.739 |
| 6% | 21.458 | 0.786 |
| 10% | 20.647 | 0.799 |
| 15% | 20.381 | 0.811 |
| 20% | 20.323 | 0.817 |
| 25% | 20.180 | 0.812 |
| 30% | 19.203 | 0.789 |
| 35% | 18.934 | 0.772 |

The impact of varying $N$ on imaging performance when $M$ is fixed at 20, 30, 40 is shown in Fig. 2(a). Similarly, Figure 2(b) demonstrates the influence of varying $M$ on imaging quality when $N$ is fixed at 20, 25, 30, 35, 40. The trend lines in these figures reveal that the SSIM values are optimized for $M$ and $N$. The SSIM values for $N$ exhibit a more pronounced decline compared to the $M$ values after reaching the peak. These results highlight the critical role of parameter selection in achieving optimal image reconstruction quality in LGSI systems.

To evaluate the LGSI imaging capabilities, image reconstruction of the resolution test target was performed at a low sampling ratio of 3 %. The various reconstruction results are illustrated in Fig. 3(a). While unsharp artifacts are evident in peripheral region of the LGSI reconstruction, their impact is significantly diminished in the central region. This phenomenon is attributed to the reduced low-order LG patterns, which result in a reduced number of features along the outer rim of the patterns.

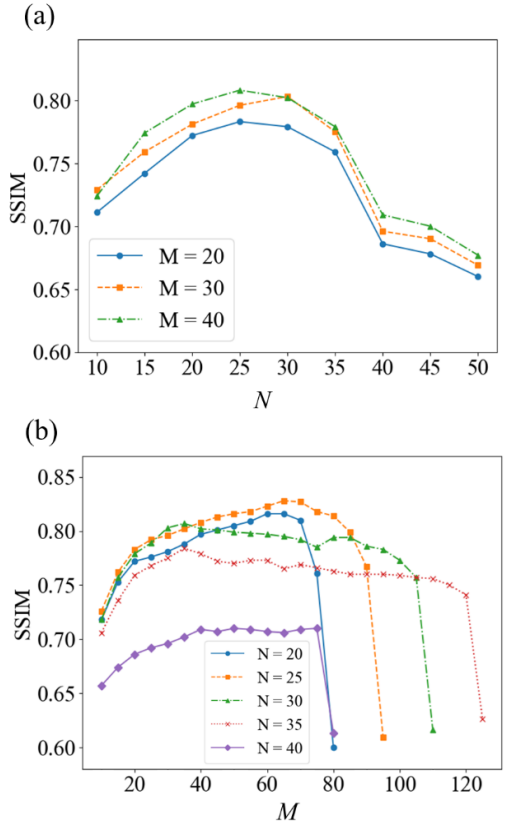

Fig. 2. (a) The SSIM values for the LG pattern are the for different orders when $M$ = 20, 30, 40, (b) different orders when $N$ = 20, 25, 30, 35, 40.

The superiority of LGSI over HSI and FSI at very low sampling ratios is demonstrated in Fig. 3(b), where a portion of the target image is enlarged for a clarity. Experimental results further corroborate this superiority (see Supplementary information Section3). It's evident that only LGSI reconstruction exhibits a high degree of fidelity in this region of the image, enabling the clear discernment of the three stripes. This advantage is also supported by additional results in supplement information Section2.

Figure 3(d) shows 480 leaves of the barrel detection signals corresponding to the Hadamard, Fourier, and LG patterns, respectively. To provide a more accurately comparison of image quality, the pixel values of the three reconstruction results are further compared by plotting against the horizontal coordinates of the reference image, as shown in Fig. 3(c). It can be seen that only the LGSI correctly shows the characteristics of the reference curve, and the pixel values of the HSI and FSI do not match the characteristics of the distribution of the reference, the curve of HSI shows erroneous peaks, the curve of FSI has only a smooth peak, further underscoring the superior performance of LGSI in preserving image fidelity at low sampling ratios.

The weight values for HSI, FSI, and LGSI, were calculated as the inner product values between the target recovery image and the patterns. These weight values were standardized into one-dimensional values with a mean of 0 and a variance of 1, respectively. The weights values of the patterns at the front of the Hadamard and Fourier patterns sequence will be much larger than that of the patterns at the backward of the sequence, because the natural image is mainly based on the low spectral information of HSI and FSI. On the other hand, in the LG imaging method, the distribution of the weight values is more uniform, which to some extent reflects the superiority of the LG's higher-order patterns utilization efficiency to the extent of image reconstruction at a very low sampling ratio.

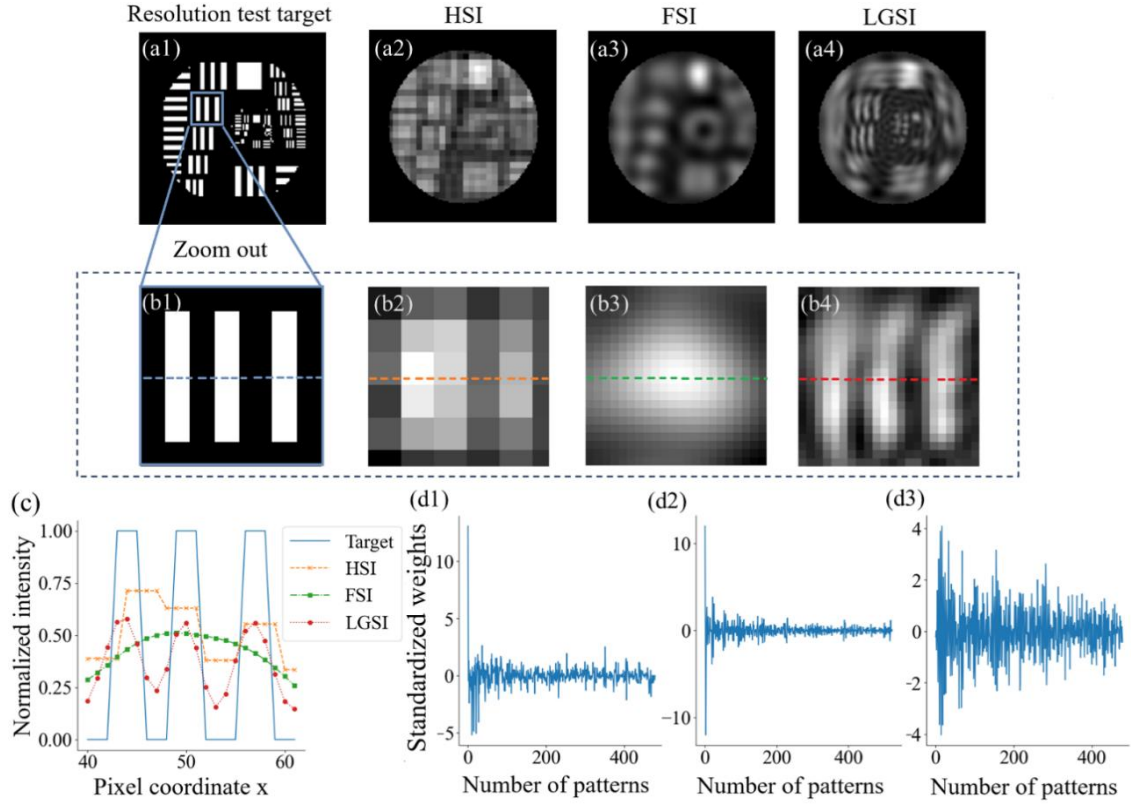

Fig. 3. (a1) Ground truth. (a2)-(a4) The reconstructed results with HSI, FSI and LGSI using 480 patterns. (b1) Local enlarge areas of (a1). (b2)-(b4) Local enlarge areas of image in (a2)-(a4). (c) Standardized intensities of corresponding line profiles in (a). The weights of Hadamard [(d1)], Fourier [(d2)] and LG [(d3)] patterns in Resolution test target.

Compared to the Fourier pattern, which rely on a single trigonometric angle with constant, LG patterns incorporate two parameters for coding the angle and the radius. Mathematically, the coefficient of the trigonometric function in the LGSI method transitions from a constant (as in FSI) to a function, introducing greater flexibility and dimensionality in pattern coding. This increased dimensionality in the base pattern coding of LGSI highlights it superiority and indicates that the LG patterns possess stronger feature representation. The results demonstrate that LGSI achieves more uniform weight distributions and superior image reconstruction quality, particularly at low sampling ratios, validating its enhanced feature representation and coding efficiency.

## Results and discussion

To experimentally validate the efficacy of the proposed SPI system, an optical experimental setup was established (see Supplementary information Section3). The experimental system employed transmissive sheet patterns of NCU logo and USAF resolution test chart as target objects. The illumination modulation pattern was set to a pixel resolution of 128×128. To demonstrate the effectiveness of LGSI imaging at low sampling ratios, a sequence of LG patterns was generated for object illumination. The reconstructed results are shown in Fig. 4. Furthermore, the reconstruction capacity of LGSI was compared with that of HSI and FSI with the results of whose are shown in Fig. 5, highlighting the superior imaging quality and efficiency of LGSI, particularly under low sampling conditions.

As demonstrated in Fig. 4, the achieved reconstructions using LGSI maintain robust imaging capacity even at reduced sampling ratios. With increasing sampling ratios, the clarity of the reconstructed images improves, as showed in Fig. 4(a)-(b). At sampling ratios 15% and 20%, more pronounced fluctuations are observed, indicating enhanced contrast in the reconstruction results. However, the characteristic features of the LG patterns become discernible and induce background noise, as can be seen in Fig. 4(c).

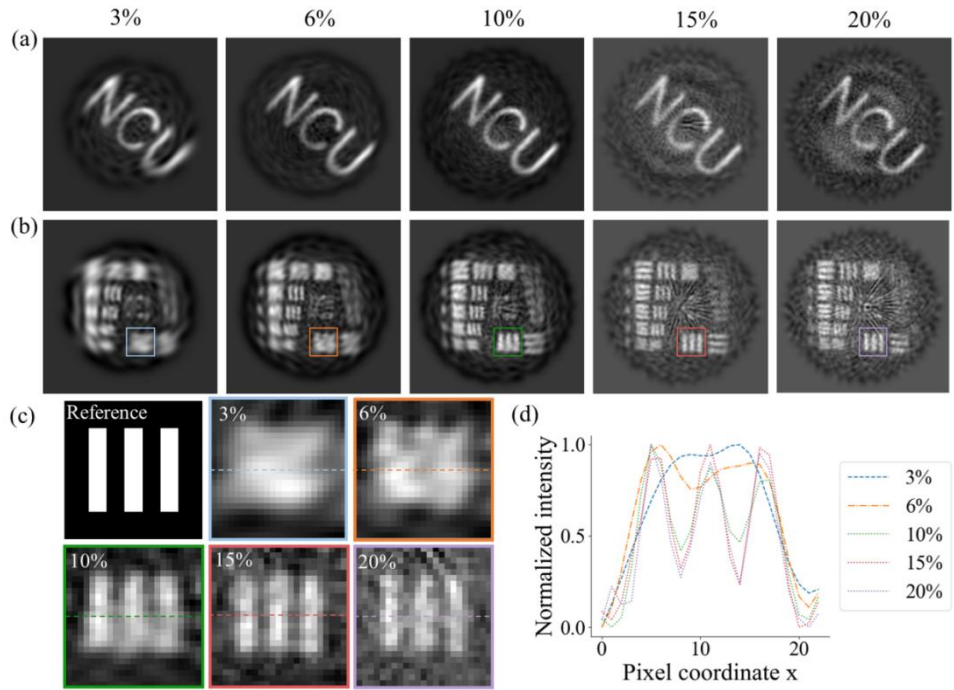

Fig. 4. Experimental reconstructed images of (a) NCU and (b) USAF resolution test chart at different sampling ratios with LGSI method. (c) Comparison of local enlarge imaging areas enclosed in (b). (d) Normalized intensities of corresponding line profiles in (c).

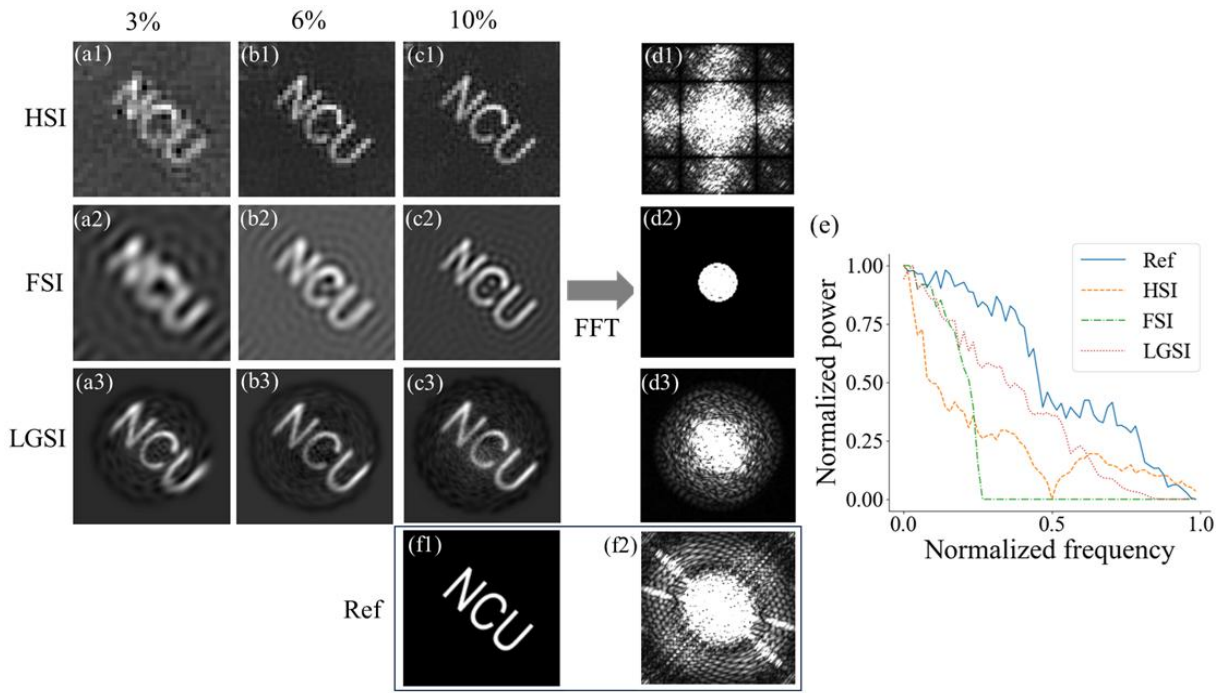

Fig. 5. Reconstructions of NCU with HSI, FSI and LGSI. (a1)- (c3) Reconstructions at different sampling ratios with three methods. (d1)- (d3) The Fourier spectrum images corresponding of (c1)-(c3). (e) Normalized power spectrum curves of (d1)-(d4). (f1) Original image of object and (f2) the corresponding Fourier spectrum.

Figure. 5(a1)-(a3) highlights the superior image clarity of LGSI compared to HSI and FSI, particularly at lower sampling ratios. The discernible contour in Fig. 5(a3) stands in stark contrast to the indistinct contours observed in Fig. 5(a1) and (a2). As demonstrated in Fig. 5(c1)-(c3), at the same measurement levels, LGSI consistently delivers more accurate reconstruction outcomes. Notably, LGSI eliminates the mosaic effect commonly associated with HSI and effectively suppresses artefactual noise in the imaging area compared to FSI.

The Fourier transformed images corresponding to Fig. 5(c1)-(c3) are shown in Fig. 5(d1)-(d3), respectively, while Fig. 5(e) displays the normalized power spectrum curves. These results reveal distinct differences among the methods, with the LGSI curve being closest to the reference. It is particularly evident that FSI reconstructs images at low frequencies, thereby validating the assertion of a paucity of high-frequency information, as evidenced by Fourier spectrograms and power spectrum curves.

## Conclusions

We have proposed and experimentally validated a physically enhanced high-quality and low-sampling-ratio SPI technique by utilizing the orthogonal mechanism of physical LG moments. Unlike conventional HSI and FSI, where imaging performances is highly sensitive to sampling strategies and optimization algorithms, our LGSI provides a facile and optimization-free approach to developing novel and high-performance imaging toolkit. LGSI achieves exceptional image quality with an extremely low sampling ratio (3%), as evidenced by a SSIM of 0.739 and a PSNR of 20.762dB. Both the experimental and simulation results demonstrate excellent agreement, further validating the robustness of the proposed method.

To substantiate the imaging capability of LGSI method, the calculated power spectra were compared with those of HSI and FSI under identical low sampling condition. The results confirm the superiority of LGSI in capture high-frequency information and achieving higher fidelity in image reconstruction. Furthermore, the weight distribution of LGSI is smoother than that of the HSI and FSI, indicating more efficient utilization of the encoding pattern information. The two-dimensions coding of the LG pattern along radial and angular directions provides a higher dimensionality in pattern representation, further enhancing the method's effectiveness. The proposed LGSI technique enables high-quality image reconstruction through a single-pixel detector, making it a promising solution for imaging applications in weak-intensity and non-visible light imaging scenarios. Moreover, the spatially structured modulation enabled by physical LG moments holds potential for applications in complex amplitude SPI, diffraction free high-quality imaging of complicated scenes, and multi-dimensional light manipulation integrated with other advanced optical techniques. These capabilities position LGSI as a versatile and practical tool for next generation imaging systems.